\documentstyle[multicol,twoside,eqsecnum,aps,pre]{revtex}
\makeatletter

  \def\refrule{%
  \end{multicols}\widetext\vglue13pt %
  \hskip10.25pc\rule{20pc}{.1mm}\hfill %
  \vglue.14cm\begin{multicols}{2}\narrowtext %
  }

  \def\references{%
  \list{\@biblabel{\arabic{enumiv}}}%
  {\labelwidth\WidestRefLabelThusFar  \labelsep2pt %
  \leftmargin\labelwidth %
  \advance\leftmargin\labelsep %
  \ifdim\baselinestretch pt>1 pt %
  \parsep  4pt\relax %
  \else   %
  \parsep  0pt\relax %
  \fi
  \itemsep\parsep %
  \usecounter{enumiv}%
  \let\p@enumiv\@empty
  \def\theenumiv{\arabic{enumiv}}%
  }%
  \let\newblock\relax %
  \sloppy\clubpenalty4000\widowpenalty4000
  \sfcode`\.=1000\relax
  \ifpreprintsty\else\small\fi
  }

  \newfont{\rmtop}{cmr10 at 9pt}
 
  \def\AUTHORS{\rmtop IGL\'OI, TURBAN, AND RIEGER}
  \def\JOURNAL{\ }
  \def\DATE{SEPTEMBER 1998}
  \def\TITLE{\rmtop ANOMALOUS DIFFUSION IN APERIODIC ENVIRONMENTS}
  \def\VOLUME{\ }
  \def\NUMBER{\ }
  \def\PAGE{1}

  \def\ps@plain{%
  \gdef\@oddhead{\ifnum\thepage=\PAGE {\hbox to 2in{\rmtop\JOURNAL%
  \hfil}\hfil{{\rmtop \VOLUME \NUMBER}}\hfil\hbox to %
  2in{\hfil{\rmtop\DATE}}}\else{\hbox to %
  1in{${\hbox{\rmtop\VOLUME}}$\hfil}\hfil\TITLE\hfil\hbox to %
  1in{\hfil\rmtop\thepage}}\fi}%
  \gdef\@evenhead{\hbox to %
  1in{\rmtop\thepage\hfil}\hfil\AUTHORS%
  \hfil\hbox to 1in{\hfil${\hbox{\rmtop\VOLUME}}$}}%
  \gdef\@oddfoot{\ifnum\thepage=\PAGE{\hbox to %
  2.5in{\hfil}\hfil$\ \ {\hbox{\rmtop\VOLUME}}\ \ \ \ \ \ %
  ${\rmtop\thepage}%
  \hfil\hbox to 2.5in{\hfil Typeset using REV\TeX}}\else{}\fi}%
  \gdef\@evenfoot{{}}%
  }%

\makeatother

\begin{document}

\title{Anomalous Diffusion in Aperiodic Environments}

\author{Ferenc Igl\'oi}
\address{Research Institute for Solid State Physics and Optics,
H-1525 Budapest, P.O.Box 49, Hungary$^*$\\
Laboratoire de Physique des Mat\'eriaux, Universit\'e Henri Poincar\'e 
(Nancy 1), F-54506 Vand\oe uvre l\`es Nancy, France }

\author{Lo\"{\i}c Turban}
\address{Laboratoire de Physique des Mat\'eriaux, Universit\'e Henri 
Poincar\'e (Nancy 1), F-54506 Vand\oe uvre l\`es Nancy, France }

\author{Heiko Rieger}
\address{HLRZ, Forschungszentrum J\"ulich, 52425 J\"ulich, Germany}

\date{1 September 1998}

\maketitle

\begin{abstract}
  We study the Brownian motion of a classical particle in
  one-dimensional inhomogeneous environments where the transition
  probabilities follow quasiperiodic or aperiodic distributions.
  Exploiting an exact correspondence with the transverse-field Ising
  model with inhomogeneous couplings we obtain many new analytical
  results for the random walk problem. In the absence of global bias
  the qualitative behavior of the diffusive motion of the particle and
  the corresponding persistence probability strongly depend on the
  fluctuation properties of the environment. In environments with
  bounded fluctuations the particle shows normal diffusive motion and
  the diffusion constant is simply related to the persistence probability.
  On the other hand in a medium with unbounded fluctuations
  the diffusion is ultra-slow, the displacement of the particle grows
  on logarithmic time scales. For the borderline situation with
  marginal fluctuations both the diffusion exponent and the
  persistence exponent are continuously varying functions of the
  aperiodicity. Extensions of the results to disordered
  media and to higher dimensions are also discussed.
\end{abstract}

\vglue.8truecm

\begin{multicols}{2}
\narrowtext

\section{Introduction}

The Brownian motion is perhaps the best understood stochastic
process in classical physics, both in homogeneous environments and  
disordered media. The study of the diffusion problem in inhomogeneous
environments is physically motivated by transport processes (molecular
diffusion, flow lines in a porous medium, electrical conduction) on one  
hand and the relaxational properties of disordered systems (random magnets,
spin-glasses) on the other (see Refs.~\cite{alexander81,bouchaud90}).

In the presence
of asymmetric transition rates, i.e., when the the probability per unit time
$w_{{\bf r},{\bf r'}}$ for a particle to jump from site ${\bf r}$ to site 
$\bf r'$ is different from $w_{{\bf r'},{\bf r}}$, the disorder strongly
modifies the behavior of the diffusive motion in $d<2$ dimensions. In one
space dimension, where the effect of disorder is most pronounced, the
diffusion is ultra-slow and the averaged mean-square displacement grows on a
logarithmic time scale~\cite{sinai82}:
\begin{equation}  
[\langle X^2(t)\rangle]_{av}\sim\ln^4 t\,.
\label{sinai}
\end{equation}  
Another type of inhomogeneity is provided by fractal lattices, either regular
or random, such as  percolation clusters, in which the Brownian motion
has been intensively studied under the name ``the ant in the
labyrinth"~\cite{degennes76,havlin87,bouchaud90}. Also much work has been
devoted
to the clarification of diffusion processes in the presence of hierarchically
distributed energy barriers~\cite{havlin87,giacometti91}, a problem which is
related to relaxation processes in disordered systems~\cite{relax}.

In the present work we study the Brownian motion in
inhomogeneous environments where the transition rates are asymmetric
and  distributed according to quasiperiodic or, more generally,
aperiodic rules. As a related earlier work, one may mention an investigation
of the Brownian motion on the two-dimensional Penrose lattice, where normal
diffusive behavior has been found~\cite{langie92}. Here we mainly concentrate
on one-dimensional aperiodic systems. Besides its mathematical interest, the
present study is also physically motivated since artificial multilayer systems
with controlled distributions of the atomic layers may now be grown by
molecular beam epitaxy~\cite{multilayers}. When particle transport has
different time scales for the motion parallel and perpendicular to the layers,
respectively, a one-dimensional diffusion process perpendicular to the
layers is, in principle, observable.

The study of cooperative phenomena in quasiperiodic and aperiodic
systems is an intensive field of research. One may mention phase
transitions and critical phenomena in Ising and other magnetic models,
percolation, self-avoiding walks, etc. Aperiodic structures, which
interpolate between periodic and random
systems, may or may not influence qualitatively the properties of a
cooperative process. Concerning the critical behavior of aperiodic
magnetic systems, a relevance-irrelevance criterion has been
proposed~\cite{luck93a,rel-crit-aper}, which is an extension of the well-known
Harris criterion for disordered systems~\cite{harris74}.  The vast amount of
exact results about the critical properties of aperiodic quantum Ising
chains~\cite{turban94,igloi94,igloiturban96,itksz,grimmbaake} and
related aperiodically layered two-dimensional Ising models~\cite{igloilajko}
are all in
accordance with this criterion. Aperiodicity may also change
to second order a transition which is of the first order in the pure system, 
as was demonstrated recently in a numerical study~\cite{berche98}.

In the present work we show that a relevance-irrelevance criterion, similar to
that of magnetic systems, can be formulated for the Brownian motion, which is
then checked against exact results obtained for different one-dimensional
aperiodic environments.

The paper is organized as follows. In Section II we introduce the basic
notations and quantities (drift velocity, diffusion constant,
persistence probability) and present the relevance-irrelevance
criterion for one-dimensional aperiodic environments. In Section III an exact
correspondence between the random walk (RW) and the transverse-field
Ising model (TIM) is presented in one dimension, which is then used in
Section IV to obtain analytical results for irrelevant, relevant and
marginal aperiodic environments. Our results are extended to higher dimensions
in Section V and discussed in the final section.

\section{Formalism and relevance-irrelevance criterion}

We consider a one-dimensional RW with nearest neighbour hopping,
characterized by transition probabilities $w_{i,i\pm 1}$
for a jump from site $i$ to site $i\pm 1$. The time evolution of $P_i(t)$,
the probability for the particle to be on site $i$ at time $t$, is governed by
the Master equation:
\begin{equation}  
{d P_i\over d t}=w_{i-1,i} P_{i-1}-(w_{i,i-1}+w_{i,i+1})P_i
+w_{i+1,i} P_{i+1}\,.
\label{master}
\end{equation}  
The transition probabilities are generally nonsymmetric, here we suppose that
their ratio is given by:
\begin{equation}  
{w_{i,i+1}\over w_{i+1,i}}=\epsilon_i=\epsilon R^{f_i}\,, 
\label{ratio}
\end{equation}  
where $R>0$ is the amplitude of the inhomogeneity ($R=1$ corresponds to
the homogeneous lattice) and the integers $f_i$ are taken from an
aperiodic or quasiperiodic sequence. For the sake of simplicity in the
following we take $w_{i+1,i}=w_{\leftarrow}=1$.

The aperiodic chain may be replaced by a periodic
approximant of period $N$, such that $\epsilon_i=
\epsilon_{i+N}$~\cite{derrida83,luck93a}. The aperiodic system is recovered in
the limit $N\to\infty$. Exact expressions for the drift velocity $v_d$
and the diffusion constant $D$ obtained by Derrida for an arbitrary
distribution
of the transition rates~\cite{derrida83} can then be used to treat the
aperiodic
case.

With our notations, the drift velocity of the aperiodic model may be
written as:
\begin{equation}  
v_d={1-\prod_{n=1}^N\epsilon_n^{-1}\over 1-\prod_{n=1}^N\epsilon_n^{-1} +
{1\over N}\sum_{n=1}^N\sum_{i=1}^N\epsilon_{n+1}^{-1}\epsilon_{n+2}^{-1}
\dots\epsilon_{n+i}^{-1}}\,.
\label{vd}
\end{equation}  
Defining the control parameter as
\begin{equation}  
\delta={1\over N}\sum_{n=1}^N\ln\epsilon_n\,,
\label{delta}
\end{equation}  
the drift velocity is zero for $\delta=0$, whereas for a small bias, such
that $\delta N\ll 1$, the drift velocity is proportional to $\delta$:
\begin{equation}  
v_d=\delta D_0\,,\quad {1\over D_0}={1\over N^2}\sum_{n=1}^N
\sum_{i=1}^N\epsilon_{n+1}^{-1}\epsilon_{n+2}^{-1}\dots\epsilon_{n+i}^{-1}\,.
\label{D0}
\end{equation}  
One can similarly calculate the diffusion constant $D$, which in the zero
bias case, $\delta=0$, is simply given by:
\begin{equation}  
D(\delta=0)=D_0\,.
\label{diff}
\end{equation}  
Before going to analyze the diffusive behavior of aperiodic walks we
consider another quantity of interest, the persistence probability
$P_{per}(L,t)$, which is the probability that the walker starting at
site $i=1$ does not cross its starting position until time $t$. Here
the length scale $L$ in the definition is set by the presence of an
adsorbing site at $i=L+1$, thus $w_{L+1,L}=0$. Due to this adsorbing
site the persistence probability has a finite long time limit,
$\lim_{t\to\infty}P_{per}(L,t)=p_{per}(L)$, which can be expressed
as~\cite{pre98}:
\begin{equation}  
p_{per}(L)=\left(1+\sum_{i=1}^L\prod_{j=1}^i\epsilon_j^{-1}\right)^{-1}\,.
\label{pers}
\end{equation}  
It is easy to see that, in the thermodynamic limit, $\lim_{L\to\infty}
p_{per}(L)=p_{per}$ plays the r\^ole of an order parameter: It is nonvanishing
for $\delta>0$ only. For the homogeneous system
with $R=1$ in~(\ref{ratio}):
\begin{equation}  
p_{per}^{hom}=1-\epsilon^{-1}\simeq\delta\,,
\label{pershom}
\end{equation}  
whereas at the critical point in a finite homogeneous system:
\begin{equation}
p_{per}^{hom}(L,\delta=0)={1\over L+1}\,.
\label{pershom2}
\end{equation}
$P_{per}(t)=\lim_{L\to\infty}P_{per}(L,t)$ is the usual persistence
probability for an infinite system.

Next we analyze the expressions of the basic quantities in Eqs.~(\ref{D0})
and~(\ref{pers}). The qualitative behavior for any {\it periodic system},
i.e., when $N$ is finite, is the same in the vicinity of the critical
point: The diffusion constant $D_0$ in Eq.~(\ref{D0}) is finite and the
persistence probability in~(\ref{pers}) has a linear $\delta$- or
$L^{-1}$-dependence.

For aperiodic systems, which are obtained in the limit $N\to\infty$, one has 
to consider the products of the transition rates
${\cal P}_n(L)=\prod_{i=1}^L\epsilon_{n+i}=\exp[\Delta_n(L)]$, where
$\Delta_n(L)=\sum_{i=1}^L \ln \epsilon_{n+i}$ measures the fluctuations
in the ``energy landscape" when $\ln \epsilon_i$ is related to the heigth
of the $i$-th energy barrier in an activated diffusion process. Here we 
differentiate between three possibilities:

i) For {\it bounded fluctuations} in the transition rates, i.e., with a
flat energy landscape, the products ${\cal P}_n(L)$ stay
finite at the critical point, thus the diffusion constant $D_0$ remains 
finite and the linear $\delta$-dependence of the persistence probability is
maintained. Thus this type of aperiodicity is {\it irrelevant} and the
aperiodic Brownian motion keeps the same critical properties as the 
homogeneous one.

ii) For {\it unbounded fluctuations} in the
transition rates, when the energy landscape is rough, so that
$\Delta_n(L) \sim L^{\Omega}$ with a wandering exponent $\Omega>0$,
the products ${\cal P}_n(L)$ which
appear in Eqs.~(\ref{D0}) and~(\ref{pers}) are divergent at the critical 
point when $L\to\infty$. Consequently, the diffusion constant $D_0$ vanishes
and the persistence probability displays a nonlinear $\delta$-dependence. We
conclude that an aperiodic environment with unbounded fluctuations is a
{\it relevant perturbation} for the Brownian motion. As we show later,
diffusion
in a relevant aperiodic environment is {\it ultraslow}, i.e., the mean-square
displacement grows on a logarithmic time-scale, as for the Sinai model in
Eq.~(\ref{sinai}).

iii) In the borderline case, the wandering exponent of the
environment is $\Omega=0$, thus fluctuations in the energy landscape
grow logarithmically. In this {\it marginal situation} the
products of the transition rates ${\cal P}_n(L)$ have a power low
$L$-dependence and the different physical quantities also
display power law singularities.
For this type of {\it anomalous diffusion} the drift velocity depends on the
bias as:
\begin{equation}  
v_d\sim\delta^{\tau}\,,\quad |\delta|\ll 1\,,
\label{tau}
\end{equation}  
with $\tau>1$, whereas the mean square displacement grows like
\begin{equation}  
\langle X^2(t)\rangle\sim t^{\psi}\,,
\label{psi}
\end{equation}  
with $\psi<1$.
Finally, the persistence probability is in general characterized by
algebraic singularities:
\begin{equation}  
p_{per}\sim\delta^{\chi}\,,\quad P_{per}(t)\sim t^{-\Theta}\,.
\label{beta}
\end{equation}  
The critical exponents defined above are not independent. Using results of
the next section in Eqs.~(\ref{corresp}) and~(\ref{lambdapert}), one can show
that the relevant time scale $t$ and the bias $\delta$ are related by
$t^{-1} \sim v_d \delta \sim \delta^{1+\tau}$. For the length scale, one has 
$L \sim v_d t \sim \delta^{-1}$. These relations lead to
the scaling laws
\begin{equation}  
\tau={2\over\psi} -1\,,\quad \chi={2\over\psi}\,\Theta\,,
\label{screlations}
\end{equation}  
which evidently hold for normal diffusion with $\tau=\psi=\chi=1$ and
$\Theta=1/2$. Thus two critical exponents are enough to describe the
behavior of the anomalous diffusion.

To obtain the characteristic quantities for relevant and marginal
environments one has
to solve the Master equation in~(\ref{master}), which amounts to
solve the eigenvalue problem for the transition
matrix or Focker-Planck (FP) operator,
\begin{equation}  
\underline{\underline{M}}\,\underline{v}_k=\lambda_k\underline{v}_k\,,
\quad\underline{u}_k^T\,\underline{\underline{M}} =\underline{u}_k^T
\lambda_k\,,
\label{eigenprobl}
\end{equation}  
where the matrix elements take the form
$(\underline{\underline{M}})_{i,j}=w_{i,j}$
for $i\ne j$ and $(\underline{\underline{M}})_{i,i}=-\sum_j w_{i,j}$.
All the physical properties of the model can be expressed in terms of
the left and right eigenvectors $\underline{u}_k$ and $\underline{v}_k$,
respectively, and the eigenvalues $\lambda_k$, which are nonpositive. For
example, the probability $P_{i,j}(t)$ that the walker starting at $t=0$ at
site $i$ arrives on site $j$ at time $t$ is given by:
\begin{equation}  
P_{i,j}(t)=\sum_k u_k(i) v_k(j)\exp(\lambda_k t)\,.
\label{probability}
\end{equation}  
The relevant time-scale of the problem is related to the inverse of the
largest nonzero eigenvalues and the dynamical properties of the RW
are connected to the scaling behavior of the eigenvalues of the
FP operator at the top of the spectrum. Considering a large finite system
of size $L$, under a change of the length scale by a factor of $b>1$, such
that $L'=L/b$, the eigenvalues at the top of the spectrum with $k\ll L$
are expected to transform as:
\begin{equation}  
\lambda_k'=b^{y_{\lambda}}\lambda_k\,,
\label{gap}
\end{equation}  
where $y_{\lambda}$ is the scaling dimension of the gap. From
Eq.~(\ref{gap}), the finite-size behavior of the eigenvalues
$\lambda_k(L)\sim L^{-y_{\lambda}}$ follows, thus the time and
length scales are related through $t\sim L^{y_{\lambda}}$. On the
other hand from~(\ref{psi}) $L^2\sim t^{\psi}$, thus the diffusion
exponent is given by:
\begin{equation}  
\psi={2\over y_{\lambda}}\,.
\label{psi1}
\end{equation}  
For relevant aperiodic environments, the leading eigenvalues have a stretched
exponential finite-size dependence, thus the diffusion exponent is formally
zero.

In the Master equation formalism, the persistence probability
$P_{per}(L,t)$ can be calculated by putting adsorbing sites at $i=0$
and $i=L+1$. Then $P_{per}(L,t)=P_{1,L+1}(t)$, which in the large-$t$-limit 
is just the first component of the zero-mode left eigenvector
$p_{per}(L)=u_1(1)$, as given by Eq.~(\ref{pers}).
\vglue3mm
\section{Correspondence between the random
walk and the Ising model in a transverse field}

First we rewrite the eigenvalue problem of the FP operator in
(\ref{eigenprobl}) in terms of the components of the right eigenvector
$v_k(i)$ as:
\end{multicols}
\widetext
\begin{equation}  
w_{i-1,i}\, v_{k}(i-1)-(w_{i,i-1}+w_{i,i+1})\, v_{k}(i)
+w_{i+1,i}\, v_{k}(i+1)=\lambda_k\, v_{k}(i)\,,
\label{master1}
\end{equation}  
and consider a finite system of size $L$, i.e., we put $w_{0,1}=w_{L+1,L}=0$.
Then we introduce the new variables:
\begin{equation}
v(i)=\alpha_i{\tilde v}(i)\,,\qquad
\alpha_{i+1}=\alpha_{i}\left({w_{i,i+1}\over w_{i+1,i}}\right)^{1/2}\!\!\!=
\alpha_{1}\left(\prod_{j=1}^i{w_{j,j+1}\over w_{j+1,j}}\right)^{1/2}\,,
\label{vi}
\end{equation}
in terms of which the eigenvalue problem is transformed into:
\begin{equation}  
\left(w_{i-1,i} w_{i,i-1}\right)^{1/2} {\tilde v}_{k}(i-1)
-(w_{i,i-1}+w_{i,i+1}){\tilde v}_{k}(i)
+\left(w_{i+1,i} w_{i,i+1}\right)^{1/2}{\tilde v}_{k}(i+1)=\lambda_k {\tilde
v}_{k}(i)\,,
\label{master2}
\end{equation}  
\hfill\rule[-2mm]{.1mm}{2mm}\rule{20.5pc}{.1mm}
\begin{multicols}{2} 
\narrowtext
\noindent which corresponds to a real symmetric eigenvalue problem $\sum_j
T_{ij}\tilde{v}_{k}(j)=\lambda\tilde{v}_{k}(i)$ with $T_{ij}=T_{ji}$. 
Consequently the eigenvalues $\lambda_k$ of the FP operator are real.

Next we show that the eigenvalue problem in~(\ref{master2}) formally appears
in the free-fermion representation of the transverse-field Ising spin chain,
with couplings $J_i$ and transverse fields $h_i$, described by the 
Hamiltonian: 
\begin{equation}  
H=-\sum_{i=1}^{L-1} J_i\sigma^x_i\sigma^x_{i+1} -\sum_{i=1}^L h_i
\sigma^z_i\,,
\label{hamilton}
\end{equation}  
where $\sigma^x_i$ and $\sigma^z_i$, are Pauli matrices at site $i$. The
Hamiltonian $H$ can be transformed into a free-fermion model by standard
techniques~\cite{liebetal}:
\begin{equation}  
H=\sum_k\Lambda_k (\eta_k^{+}\eta_k -1/2)\,.
\label{freefermion}
\end{equation}  
Here the $\eta_k^{+}$ ($\eta_k$) are fermion creation (annihilation)
operators and the excitation energy $\Lambda_k$ is the solution of the
eigenvalue equation:
\end{multicols}
\widetext
\noindent\rule{20.5pc}{.1mm}\rule{.1mm}{2mm}\hfill
\begin{equation}  
J_{i-1}h_{i-1}\Phi_{k}(i-1) + (J_{i-1}^2+h_i^2)\Phi_{k}(i) +
 J_i h_i\Phi_{k}(i+1)=\Lambda_k^2\Phi_{k}(i)\,.
\label{eigenising}
\end{equation}  
\hfill\rule[-2mm]{.1mm}{2mm}\rule{20.5pc}{.1mm}
\begin{multicols}{2} 
\narrowtext
\noindent Comparing Eqs.~(\ref{eigenising}) and~(\ref{master2}), one can
notice that they can be cast into the same form with the
correspondences:
\begin{equation}  
\begin{array}{rcl}
J_i &\iff &\left(w_{i+1,i}\right)^{1/2}\\
h_i &\iff &\left(w_{i,i+1}\right)^{1/2}\\
\Phi_{k}(i) &\iff  & (-1)^i {\tilde v}_{k}(i)\\
\Lambda_k^2 &\iff & -\lambda_k\,.
\end{array}
\label{corresp}
\end{equation}  
Thus there is a {\it mathematical equivalence} between the RW
in an inhomogeneous environment and the TIM with the corresponding
inhomogeneous couplings, as described in~(\ref{corresp}). One can 
however go further and show that there are several {\it physical
quantities} which are closely related in the two problems.

First let us consider the persistence probability $p_{per}(L)$, which
is calculated with adsorbing boundary conditions $w_{0,1}=w_{L+1,L}=0$
as $p_{per}(L)=u_1(1) v_1(L+1)={\tilde u}_1(1) \alpha_1^{-1} \alpha_{L+1}
{\tilde v}_{1}(L+1)=[{\tilde v}_{1}(L+1)]^2$.
Now using the correspondences in~(\ref{corresp}),
the equivalent TIM with $h_0=J_L=0$ has a fixed spin at $i=0$, whereas the
other
end of the chain at $i=L+1$ is free~\cite{igloi98}. The surface magnetization
of the chain measured on the $i=L+1$-th spin is given by 
${\overline  m}_s(L)=\Phi_{1}(L+1)$~\cite{igloi98,peschel84}, thus
using~(\ref{corresp}) we have a
relation between the surface magnetization of the TIM and the persistence
probability of the RW:
\begin{equation}  
[{\overline m}_s(L)]^2\iff p_{per}(L)\,.
\label{corresppr}
\end{equation}  
We note that this relation has already been mentioned in Ref.~\cite{pre98}.

Let us now analyze the
relation between the energy scales in the two problems, $\lambda_k$ and
$\Lambda_k$, respectively. First one may notice that, according to the last
equation in~(\ref{corresp}), the eigenvalues $\lambda_k$ of the FP operator 
are nonpositive, as they should on physical grounds.

 For the {\it periodic} TIM with $J_i=h$ and
$h_i=h_{i+N}=h\epsilon_i^{1/2}$, the energy of low-lying modes, close to the
critical point $\delta=0$, are given by a perturbation 
calculation~\cite{luck93a} as: 
\begin{equation}  
\Lambda_k^2\simeq v_s^2(\delta^2+q_k^2)\,,
\label{lambdapert}
\end{equation}  
where $q_k\sim 1/N$ denotes the wavenumber corresponding to the
smallest excitation energy. Comparing the expression of the sound velocity
$v_s$
in Ref.~\cite{luck93a} to $D_0$ in Eq.~(\ref{D0}), we obtain another
correspondence:
\begin{equation}  
v_s^2\iff D_0\,.
\label{correspD0}
\end{equation}  
Taking the infinite periodic approximant limit, $N\to\infty$, we
obtain the following important result: {\it if in an environment the
diffusion is normal, i.e., the diffusion constant $D_0$ is finite, then the
phase transition of the corresponding TIM with inhomogeneous couplings is
in the Onsager universality class}. On the other hand, if the
diffusion is anomalous, i.e., the diffusion exponent is $\psi<1$, then
the phase transition of the TIM is {\it not} of Onsager type. In the marginal
case the phase transition in the TIM is characterized by an anisotropy
exponent (or dynamical exponent) $z>1$, which describes the finite
size scaling behavior of the gap $\Lambda_k\sim L^{-z}$ for $k\ll
L$.  According to Eq.~(\ref{corresp}), $z$ is related to the diffusion
exponent through:
\begin{equation}  
z\iff {1\over\psi}\,.
\label{corresppsi}
\end{equation}  
On the other hand, in the relevant case, $\Lambda_k$ has a stretched
exponential $L$-dependence, thus the dynamical exponent is formally
infinite.

Finally, we present a useful estimate for the
smallest (in absolute value) nonzero eigenvalue of the FP operator,
$\lambda_{min}$, by transforming a related result for the
TIM~\cite{itksz,igloi98}. Having a large finite chain of length $L$ with
reflecting b.c. at $i=1$ ($w_{1,0}=w_{0,1}=0$) and adsorbing b.c. at $i=L+1$
($w_{L+1,L}=0$, $w_{L,L+1}\ne 0$), then the leading finite-size behavior of
$\lambda_{min}(L)$ is connected to the product of the persistence
probabilities at the two ends of the chain as:
\begin{equation}  
\lambda_{min}(L)\simeq - p_{per}(L) {\overline p}_{per}(L)\prod_{i=1}^L
\epsilon_i^{-1}\,,
\label{lambdamin}
\end{equation}  
where the persistence probability at site $L$ is given by:
\begin{equation}  
{\overline p}_{per}(L)=\left(1+\sum_{i=1}^{L-1}\prod_{j=1}^i
\epsilon_{L-j}^{-1}\right)^{-1}\,,
\label{persr}
\end{equation}  
whereas $p_{per}(L)$ is given by Eq.~(\ref{pers}), however for $L-1$ sites.

\section{Diffusion in aperiodic environments}

Here we consider different one-dimensional aperiodic environments and
study the corresponding diffusive behavior by analytical methods. As
we mentioned before, the relevance-irrelevance of the perturbation is
connected to the fluctuation properties of the aperiodic environment.

To be more specific, we consider the behavior at criticality ($\delta=0$) of
the logarithm of the transition rates $\ln\epsilon_i$, which is related to
the heigth of the $i$-th energy barrier, $\Delta e_i$, in an activated
diffusion process. When the global bias vanishes, $\epsilon$ takes its 
critical value $\epsilon_c$ such that, according to~(\ref{delta}),
\begin{equation}  
\epsilon_c=R^{-\rho_\infty}\,,\quad\rho_{\infty}=\lim_{L\to\infty} n_L/L\,,
\quad n_L=\sum_{i=1}^L f_i\,,
\label{epsilonc}
\end{equation}  
where $\rho_\infty$ is the asymptotic density of the perturbed transition
probabilities. Then the fluctuation of the ``energy landscape" in a large
system of size $L$ is characterized by:
\begin{eqnarray}  
\Delta(L)&=&\sum_{i=1}^L\ln
\epsilon_i=n_L\ln R + L\ln\epsilon_c\nonumber\\
&=&\ln R\left(n_L - L\rho_{\infty}\right)\sim\ln R\ L^{\Omega}\,.
\label{Omega}
\end{eqnarray}  
Here $\Omega$ is the wandering exponent of the
aperiodic sequence~\cite{Queffelec} which is easily obtained for aperiodic
substitutional sequences.

Working with a finite alphabet $A$, $B$,..., such that, via subsitution, 
$A\to S(A)$, $B\to S(B)$, etc, the sequence $f_i$ in~(\ref{ratio}) is
obtained by starting with one of the letters, iterating the substitution
process and finally replacing the letters by digits (or groups of digits) 0
and 1. For a two-letter sequence, one can directly proceed with subsitutions
on 0 and 1.

The fluctuation properties of a sequence can be deduced from its
substitution matrix with entries $n_{ij}$ giving the numbers of letters
$i=A,B,...$ in $S(j)$ ($j=A,B,...$)~\cite{Queffelec}.  The wandering exponent
$\Omega$ involves the two largest eigenvalues in modulus of this matrix,
$\mu_1$ and $\mu_2$, and reads $\Omega=\ln |\mu_2|/\ln\mu_1 $.

We now consider different environments associated with specific aperiodic
sequences. Environments with bounded ($\Omega<0$), unbounded ($\Omega>0$)
and marginal ($\Omega=0$) fluctuations are treated separately.
Most of the results can be obtained by adapting the analytical
methods developed for the aperiodic TIM , using the correspondences given
in~(\ref{corresp}), (\ref{corresppr}), (\ref{correspD0})
and~(\ref{corresppsi}).

\subsection{Aperiodic environments with bounded fluctuations}

\subsubsection{Quasiperiodic (Fibonacci) environment}

Quasiperiodic lattices can be generated in several different ways,
among others by the well known cut-and-project method. Here we
use the following algebraic definition for a one-dimensional
quasiperiodic sequence:
\begin{equation}  
f_i=1+\left[{i\over\omega}\right]-\left[{i+1\over\omega}\right]\,,
\label{quasi}
\end{equation}  
where $[x]$ is the integer part of $x$ and $\omega>1$ is irrational.

The Fibonacci sequence can be generated by the substitution rule $0\to
010$, $1\to 01$~\cite{rem-fibo} starting with 0. When read from left to 
right, it corresponds to~(\ref{quasi}) with $\omega=(\sqrt{5}+1)/2$, the
golden mean.

The sequence in~(\ref{quasi}) and the corresponding quasiperiodic
environment, as defined in Eq.~(\ref{ratio}), have bounded fluctuations 
since the wandering exponent in~(\ref{Omega}) is $\Omega=-1$.
Consequently, the diffusion constant $D_0$ in~(\ref{D0}) is finite. As shown 
in Appendix~A, it can be expressed in closed form using the methods
of Ref.~\cite{luck93a} as:
\begin{equation}  
D_0=\left({\ln R\over R^{1/2}-R^{-1/2}}\right)^2\,.
\label{quasiD0}
\end{equation}  
It is interesting to note that $D_0$ does not depend on the value of the
irrational parameter $\omega$.

The persistence probability in~(\ref{pers}) can be also evaluated 
analytically using the techniques of Ref.~\cite{turban93}(see Appendix A).
For small bias there is a linear $\delta$-dependence:
\begin{equation}  
p_{per}(\delta)=R^{1-1/\omega}{\ln R\over R-1}\,\delta\,,\quad 0<\delta\ll
1\,, \label{quasipers}
\end{equation}  
where the prefactor depends on $\omega$.

For the other end of the system, except for the first digit which is 
irrelevant, the sequence, read from right to left, is given by:
\begin{equation}  
f_i=\left[{i+\omega\over\omega^2}\right]-\left[{i+\omega-1\over
\omega^2}\right]\,,
\label{quasi2}
\end{equation}  
which leads to the persistence probability:
\begin{equation}  
{\overline p}_{per}(\delta)=R^{-1+1/\omega}{\ln R\over 1-R^{-1}}\,\delta\,,
\quad 0<\delta\ll 1\,.
\label{quasipersr}
\end{equation}  

One may notice a simple relation between the diffusion constant
and the persistence probabilities,
\begin{equation}  
p_{per}(\delta)\,{\overline p}_{per}(\delta)=D_0\,\delta^2\,,
\label{D0pers}
\end{equation}  
which is valid for any value of the irrational parameter $\omega$.

\subsubsection{Thue-Morse sequence}

The binary Thue-Morse sequence~\cite{dekking83a} is generated by the
substitutions $0\to 01$ and $1\to 10$, leading to $0110100110010110\dots$
The wandering exponent is $\Omega=-\infty$ thus the sequence has bounded
fluctuations. Consequently, the diffusion constant is finite and,
as shown in Appendix A, can be
calculated along the lines of~\cite{luck93a} as:
\begin{equation}  
D_0=\left( {2\over R^{1/4}+R^{-1/4}}\right)^4\,.
\label{D0TM}
\end{equation}  
The persistence probability is linear for small bias and given by:
\begin{equation}  
p_{per}(\delta)=\left( {2\over R^{1/4}+R^{-1/4}}\right)^2\delta\,,
\label{persTM}
\end{equation}  
in agreement with the result of Ref.~\cite{turban94}, when properly 
translated.

Since the Thue-Morse sequence is reflection symmetric, the persistence
probability is the same at both ends of the system,
$p_{per}(\delta)={\overline p}_{per}(\delta)$ so that the
relation in~(\ref{D0pers}) is again fulfilled.

Analyzing the results obtained for environments with bounded fluctuations we
are led to the following conclusions: 

i) Predictions of the
relevance-irrelevance criterion are fully satisfied: The diffusion constant 
is finite and the critical exponents ($\tau=\psi=\chi=1$) take the same
values as in homogeneous environments. 

ii) In both examples, the diffusion constant $D_0$ is invariant under 
the transformation $R\to1/R$. This is in agreement with the fact that 
$D_0(R)$ is maximal for the homogeneous system, i.e., at $R=1$. 

iii) For symmetric sequences, the persistence probabilities $p_{per}$ and
${\overline p}_{per}$ are equal, whereas, for asymmetric sequences, they 
satisfy the relation $p_{per}(R)={\overline p}_{per}(1/R)$.

\subsection{Aperiodic environments with unbounded fluctuations}

The aperiodic environments display unbounded fluctuations when the
wandering exponent $\Omega>0$. Consequently, the diffusion constant
in~(\ref{D0}) vanishes in the infinite periodic approximant limit and the
persistence probability in~(\ref{pers}), like the surface magnetization in
the TIM, has an anomalous behavior.

To obtain a qualitative picture for the behavior of the RW in such an
environment, we estimate the leading eigenvalue of the FP operator
$\lambda_{min}$ in a finite system of size $L$, taking into account 
that the relevant time-scale of the process is given by 
$t\sim\lambda_{min}^{-1}$.

According to Eq.~(\ref{lambdamin}), the size dependence of
$\lambda_{min}$ for a sequence with unbounded fluctuations is
primarily determined by the product $\prod_{i=1}^{L-1}\epsilon_i^{-1}
\sim\exp(-{\rm const} L^{\Omega})$. Thus, at criticality,
the leading eigenvalue has a stretched exponential size-dependence:
\begin{equation}  
\lambda_{min}\sim\exp(-{\rm const}\ L^{\Omega})\,.
\label{lambdarel}
\end{equation}  
This implies an ultra-slow diffusion process, which takes place on a 
logarithmic time-scale:
\begin{equation}  
\langle X^2(t)\rangle\sim (\ln t)^{2/\Omega}\,.
\label{apersinai}
\end{equation}  
In a random environment, with $\Omega=1/2$, Eq.~(\ref{apersinai}) 
corresponds to the Sinai diffusion in~(\ref{sinai}).

Next, we consider the persistence probability $p_{per}$ and analyze its
properties for an exactly solvable environment with unbounded fluctuations.

\subsubsection{Rudin-Shapiro environment - persistence probability}

The Rudin-Shapiro sequence~\cite{dekking83a} is obtained via substitutions 
on pairs of digits, $00\to 0001$, $01\to 0010$, $10\to 1101$ and $11\to 1110$,
such that starting on $00$, one generates the sequence $0001001000011101\dots$
The wandering exponent of the sequence, $\Omega=1/2$, is the same as for a
random environment.

In the critical situation, the persistence probability can be exactly
calculated using the methods of~\cite{igloi94} and the correspondence in
(\ref{corresppr}). One obtains completely different behaviors for $R>1$ and
$R<1$:

i) For $R>1$, the persistence probability remains finite for an infinite
system at criticality:  %
\begin{equation}  
p_{per}=1-{R^{1/2}\over 1-R^{1/2}+R}\,,
\label{RSpers+}
\end{equation}  
thus the transition in the persistence properties is {\it discontinuous}
as the bias $\delta$ changes sign.

ii) For $R<1$, the persistence probability has an anomalous size dependence
at the critical point:
\begin{equation}  
p_{per}\sim\exp(- {\rm const}\, L^{1/2})\,,
\label{RSpers-}
\end{equation}  
which goes to zero in the infinite-size limit.

Reading the sequence from the other end, amounts to exchange the digits
$1\leftrightarrow 0$, which is equivalent to the transformation $R\to 1/R$.
Thus the persistence probability of the critical Rudin-Shapiro environment
has very different properties for the two ends of the chain: While it stays
finite at one end, it vanishes at the other. At this point, one may ask how
the persistence probability behaves if the system starts at an arbitrary
position $i$ along the sequence. One can answer by using the correspondences 
in Section III and the known results for the surface magnetization of the
Rudin-Shapiro TIM model~\cite{aperiodic}. {\it Typically}, i.e., with
probability one, the critical point persistence vanishes, as
in~(\ref{RSpers-}). However, there is a fraction of starting points, the
so-called ``{\it rare events}", $p_{rare}\sim L^{-\Theta_{av}}$, for systems 
of size $L$, where the persistence probability is finite $p_{per}=O(1)$, as
in~(\ref{RSpers+}). Then the {\it average persistence}, which is obtained by
averaging over the starting positions, scales as the probability of the rare
events, thus
\begin{equation}  
\left[p_{per}(L)\right]_{av}\sim p_{rare}(L)\sim L^{-\Theta_{av}}\,,
\label{persav}
\end{equation}  
where $[\dots]_{av}$ denotes an average over the starting position. The same
scenario remains valid for other aperiodic sequences with unbounded
fluctuations. Translating exact results for the TIM~\cite{aperiodic} into the
RW language, the persistence exponent $\Theta_{av}$ is connected to
the wandering exponent $\Omega$ of the sequence via:
\begin{equation}  
\Theta_{av}=1-\Omega\,.
\label{thetaav}
\end{equation}  
To summarize, the diffusion process in an aperiodic environment with
unbounded fluctuations is anomalous: i) The average displacement grows
logarithmically in time and ii) The persistence probability is {\it
not self-averaging}, its averaged dependence involves the
fluctuation exponent $\Omega$ of the environment.

\subsection{Aperiodic environments with marginal fluctuations}

Aperiodic environments with marginal fluctuations are characterized by
a wandering exponent $\Omega=0$. The critical behavior of the TIM with
marginally aperiodic couplings is nonuniversal and several
critical exponents vary continuously with the strength of the
aperiodic perturbation.  From the correspondences presented in Section
III, it follows that the diffusive behavior of the RW in this type of
environment is also anomalous.  Generally, the diffusion constant $D_0$
vanishes and both the diffusion exponent, $\psi=\psi(R)<1$, and the
persistence exponent, $\chi=\chi(R)>1$ are continuous functions of
the asymmetry parameter $R$.

For aperiodic environments with marginal fluctuations, the scaling
behavior of the eigenvalues of the FP operator at the top of the spectrum
can be obtained exactly by a renormalization group (RG) transformation, as
introduced for the TIM in Ref.~\cite{igloiturban96}, later applied
to different sequences in~\cite{itksz} and generalized 
in~\cite{grimmbaake}.

The essence of the method is a decimation procedure such that, after 
one step, a fraction $1/b$ of the original lattice sites are left. The 
master equation for the renormalized system can be cast into the same form
as  the original one, provided a finite set of appropriate new parameters 
are introduced into the original equation (parameter space). The linearized
transformation at the fixed point corresponding to $\lambda^*=0$ gives 
the gap exponent $y_\lambda$ as indicated in Eq.~(\ref{gap}).
To construct explicitly the RG equations, we refer to the related work on
the TIM in Refs.~\cite{igloiturban96,itksz,grimmbaake}.  
The same type of RG procedure can be used to calculate the persistence
exponent, but one can also simply deduce it from  a finite-size-scaling
analysis of Eq.~(\ref{pers}) at criticality as shown in Appendix B for
a specific sequence.

In the following we present results translated from 
Refs.~\cite{igloiturban96,itksz} for several environments
with marginal fluctuations.

\subsubsection{Period-doubling environment}

Using the substitutions $0\to 01$ and $1\to 00$ and starting
with $0$, one generates the period-doubling sequence~\cite{collet80}
$0100010101000100\dots$ which, apart from the last digit, is symmetric and 
has a vanishing wandering exponent.

The diffusion exponent in the period-doubling environment is given by:
\begin{equation}  
\psi={\ln 2\over\ln\left( R^{1/6}+R^{-1/6}\right)}\,,
\label{psipd}
\end{equation}  
whereas the persistence exponents, which are the same at both ends due to
symmetry, read:
\begin{equation}  
\chi={\overline\chi}={\ln\left( R^{1/6}+R^{-1/6}\right)\over\ln 2}\,,
\label{betapd}
\end{equation}  
as shown in Appendix B.

\subsubsection{Paper-folding environment}

The paper-folding sequence can be generated by a recurrent folding of a sheet
of paper~\cite{dekking83a}. It corresponds to the two-letter substitutions
$00\to 0010$,  $11\to 0111$, $10\to 0110$ and $01\to 0011$. Starting with 
$00$ one arrives at $0010011000110110\dots$. This environment has the same
($R\leftrightarrow1/R$) type of asymmetry as for the Rudin-Shapiro sequence,
if one forgets the last, irrelevant, digit. The wandering exponent vanishes.

The diffusion exponent in the paper-folding environment is
\begin{equation}  
\psi={\ln 2\over\ln\left( R^{1/4}+R^{-1/4}\right)}\,.
\label{psipf}
\end{equation}  
The persistence exponents are different at the two ends of the system and
are given by:
\begin{equation}  
\chi={\ln\left( 1+R^{-1/2}\right)\over\ln 2}\,,
\quad{\overline\chi}={\ln\left( 1+R^{1/2}\right)\over\ln 2}\,.
\label{betapf}
\end{equation}  

\subsubsection{Hierarchical environment}

Here we consider the Huberman-Kerszberg hierarchical
environment~\cite{hier-seq}, where the positions $i$ of the digits $f_i$
satisfy the
relation:
\begin{equation}  
i=2^{f_i}(2l+1)\,,\quad l=0,1,2,\dots\,,
\label{hier}
\end{equation}  
thus the sequence starts as $0102010301020104\dots$. The diffusion problem
in the same environment with symmetric transition rates has already been
studied before~\cite{giacometti91}.

In the nonsymmetric case the diffusion exponent is
\begin{equation}  
\psi={\ln 2\over\ln\left( R^{1/2}+R^{-1/2}\right)}\,.
\label{psihier}
\end{equation}  
The persistence exponents are different at the two ends of the system and
are given by:
\begin{equation}  
\chi={\ln\left( 1+R^{-1}\right)\over\ln 2}\,,
\quad {\overline\chi}={\ln\left( 1+R\right)\over\ln 2}\,.
\label{betahier}
\end{equation}  
One can also easily solve the diffusion problem for
generalized hierarchical environments following the solution of the
corresponding TIM in Ref.~\cite{itksz}.

To summarize, the diffusion in marginally aperiodic environments is 
anomalous, both the diffusion exponent $\psi$ and the persistence exponents
$\chi$, ${\overline\chi}$ are continuous functions of the parameter $R$, 
they however satisfy the scaling relation
\begin{equation}  
\chi(R)+{\overline\chi}(R)={2\over\psi(R)}\,,
\label{betapsi}
\end{equation}  
which is a consequence of~(\ref{lambdamin}).

\section{Relevance-irrelevance criterion in higher dimensions}

The relevance-irrelevance criterion of Section II can be generalized
for a $d$-dimensional environment, where the nonsymmetric transition
rates $w_{{\bf r},{\bf r'}}$ are perfectly correlated in $(d-D)$ dimensions.
Thus they vary in $D\le d$ dimensions and the fluctuations are characterized
by a wandering exponent $\Omega$.

In our considerations the basic r\^ole is played by the displacement
probability $P(L)$, which measures the fraction of walks that have moved to a
distance $L$ from their starting position during time $t \sim L^2$, which is
the characteristic time-scale for a homogeneous medium.
In a one-dimensional homogeneous environment with a weak uniform bias
$0<\delta_u=\ln (w_{\leftarrow}/w_{\rightarrow})\ll L^{-1}$, the displacement
probability in the unfavorable direction is $P_u(L)
\sim\exp(-{\rm const}\ \delta_u L)$, thus exponentially small.

In the absence of a global bias, an inhomogeneous
(random or aperiodic) environment does not generally favor a net
displacement of the particle. But due to the fluctuations in the
transition rates, the motion can be  effectively biased locally.
We now estimate the effective force or average local bias $\delta(L)$
inside a domain $U_L$ of linear size $L$.
First, generalizing the expression in~(\ref{Omega}), we calculate the
accumulated value of $\ln (w_{{\bf r},{\bf r'}}/w_{{\bf r'},{\bf r}})$ 
in the domain as:
\begin{equation}  
\Delta(L)=\sum_{{\bf r}\in U_L}\ln (w_{{\bf r},{\bf r'}}/w_{{\bf r'},{\bf
r}}) \sim\ln R\ L^{D\Omega}\,.
\label{OmegaD}
\end{equation}  
Then the averaged local bias along the inhomogeneous directions is given by:
\begin{equation}  
\delta(L)\sim {\Delta(L)\over L^D}\sim L^{-D(1-\Omega)}\,.
\label{deltaL}
\end{equation}  
Thus the displacement probability, in analogy with the uniform case, is
given by:
\begin{equation}  
P(L)\sim\exp[- {\rm const}\ \delta(L) L]\sim\exp\left[-{\rm const}\
L^{1-D(1-\Omega)}\right]\,.
\label{pL}
\end{equation}  
Now, depending on the sign of the exponent
\begin{equation}  
\phi=1-D(1-\Omega)\; ,
\label{phi}
\end{equation}  
the behavior of the displacement probability and, consequently, the
diffusion properties are different.

i) For $\phi<0$, i.e., for $\Omega < 1-1/D$, the displacement
probability has no exponential size dependence, thus one has the same
type of diffusive behavior as for homogeneous systems with zero bias.
Consequently, the environment does not modify the normal diffusive
motion of the particle and therefore this type of perturbation is {\it
  irrelevant}.

ii) On the other hand, for $\phi>0$, i.e., for $\Omega > 1-1/D$, the
displacement probability decays exponentially with $L^\phi$, thus this type
of environment is {\it relevant} for the diffusive properties. The
relation between the relevant time-scale $t$, which is proportional to
the characteristic number of steps needed to have a displacement $L$, and 
the length scale $L$ is obtained as:
\begin{equation}  
t\sim P(L)^{-1}\sim\exp\left({\rm const}\ L^{-D(1-\Omega)+1}\right)\,.
\label{tpL}
\end{equation}  
Thus the mean square displacement, which is proportional to $L^2$,
grows on a logarithmic scale as:
\begin{equation}  
\langle X^2(t)\rangle\sim (\ln t)^{{2\over 1-D(1-\Omega)}}\,.
\label{genersinai}
\end{equation}  
iii) The borderline case, $\phi=0$, corresponds to the marginal situation 
where the $L$-dependence in the exponential of Eq.~(\ref{pL}) can be
logarithmic, leading to nonuniversal diffusive behavior, as observed
in Section IV for one-dimensional systems.

We note that the above relevance-irrelevance criterion is in complete
agreement with the exact results we obtained above for
$d=D=1$. For example Eqs.~(\ref{apersinai}) and~(\ref{genersinai})
coincide in this case.  On the other hand, in the case of uniform
disorder with $d=D$ and $\Omega=1/2$, the borderline dimension
predicted by Eq.~(\ref{phi}) is $d^*=2$, which is in agreement with the
results of RG investigations~\cite{d*}.

\section{Discussion}

In this paper we have studied the scaling properties of the Brownian
motion in inhomogeneous environments where the transition rates are
asymmetric and their variation follows some quasiperiodic, aperiodic
or hierarchical rules. It has been shown that the diffusive motion of a
particle in such type of environments can be anomalous and a
relevance-irrelevance criterion has been formulated, which allows to predict
the different scenarios. In one dimension, we have obtained many exact 
results which all are in agreement with the abovementioned
relevance-irrelevance criterion. In these one-dimensional calculations, we
have made use of an exact mathematical correspondence between the Master
equation of the RW and the eigenvalue problem for the energy of the
free-fermionic excitations of an inhomogeneous TIM. This correspondence has
been exploited to obtain relations between different physical quantities in
the two problems. The analytical results previously obtained for the TIM have
been translated into exact results for the diffusion problem.

At this point we note that there is another model of statistical physics, 
the {\it directed walk}, which is also closely connected to the TIM.
As was shown in Ref.~\cite{igloiturban96}
the scaling properties of the directed walk are connected to the eigenvalues
of Eq.~(\ref{eigenising})
at the {\it top of the spectrum}. Thus the three problems (Ising model, 
random walk and directed walk) are inherently related, the complete solution
of any of those contains the necessary information about the properties of 
the two others. In particular, one single RG transformation describes the
scaling properties of the three models: the fixed point at $\Lambda=0$
for the TIM governs the critical properties
of the Ising model and that of the Brownian motion, whereas the fixed point
at the top of the spectrum is connected to the properties of the directed 
walk.

Next one can show that the parametrization of the transition probabilities,
$w_{i+1,i}=1$ below Eq.~(\ref{ratio}), does not affect the conclusions of 
the paper. The relevance-irrelevance criterion in Section II is evidently
unaffected by this restriction, and in the marginal situation the 
nonuniversal exponents $\chi$, $\overline{\chi}$ and $\psi$ are also
insensitive to this parametrization. For the persistence exponents it follows
from the fact that in Eqs.~(\ref{pers}) and~(\ref{persr}), only the ratio of
the transition rates appears. From Eq.~(\ref{betapsi}), or more generally
from the relation in Eq.~(\ref{lambdamin}), the same conclusion is reached
for the exponent $\psi$. This result can be also obtained by analyzing the
structure of the systematic RG-technique of Ref.~\cite{grimmbaake}.

The properties of the Brownian motion in a relevant aperiodic environment
are in many respect similar to those in a disordered media. In the critical
situation, i.e., in the absence of global bias, the diffusion is ultraslow 
in both cases and the corresponding relations in~(\ref{sinai}) and
(\ref{apersinai}) are analogous. In the off-critical situation, however,
there is an important difference between the diffusive properties in the
two environments. In a disordered media for small enough global bias,
such as $0<\delta<\delta_+$, the diffusive motion of the particle is
anomalous and the average displacement of the walker grows as a power law,
$[\langle X(t)\rangle]_{av}\sim t^{\mu}$, with a
$\delta$-dependent exponent $0<\mu(\delta)<1$~\cite{derridapomeau}. This
regime is in complete correspondence with the Griffiths-McCoy phase of the
random transverse-field Ising spin chain\cite{pre98}. As
shown very recently in Ref.~\cite{aperiodic2}, this anomalous diffusion 
regime is absent for relevant aperiodic environments.

\acknowledgements
We thank Dragi Karevski for useful discussions. This work has been 
supported by the French-Hungarian cooperation
program "Balaton" (Minist\`ere des Affaires \'Etrang\`eres-O.M.F.B.),
the Hungarian National Research
Fund under grants No OTKA TO23642, OTKA TO25139 and OTKA
TO15786 and by the Ministery of Education under grant No FKFP
0765/1997.  H.~R. was supported by the Deutsche
Forschungsgemeinschaft (DFG). The Laboratoire de Physique des Mat\'eriaux
is Unit\'e Mixte de Recherche C.N.R.S. No 7556.

\begin{appendix}
\section{Diffusion constant and persistence probability for the Fibonacci 
and Thue-Morse environments}

Using (\ref{epsilonc}), the expression for the diffusion constant $D_0$ 
in~(\ref{D0}) can be rewritten at criticality as:
\begin{eqnarray}  
{1\over D_0}&=&{1\over N^2} \sum_{n=1}^N \sum_{i=1}^N \epsilon_c^{-n-i}
R^{-\sum_{j=1}^i f_{j+n}}\nonumber\\
&=&{1\over N^2} \sum_{n=1}^N \sum_{i=1}^N
R^{g_n-g_{n+i}}\,,
\label{dif}
\end{eqnarray}  
where $g_i=n_i-i\rho_\infty$.

Similarly, the persistence probability in Eq.~(\ref{pers}) is given 
at criticality by:
\begin{equation}  
{1\over p_{per}(L)}=1+\sum_{i=1}^L R^{-g_i}\,.
\label{persl}
\end{equation}  

For the aperiodic environments, in the limit $N\to\infty$ and $L\gg1$, 
respectively, the powers of $R$ in Eqs.~(\ref{dif}) and~(\ref{persl}) 
can be replaced by their averaged values.

i) For the Fibonacci sequence, $g_i+1/\omega$ 
is the fractional part of $(i+1)/\omega$, which is uniformly distributed
over $[0,1]$ for $\omega$ irrational. Thus, in the limit $N\to\infty$,  the
average in~(\ref{dif}) can be replaced by an integral,
\begin{equation}  
{1\over D_0}=\int_0^1 R^g dg \int_0^1 R^{-g} dg\,,
\label{averD0quasi}
\end{equation}  
which leads to the result given in Eq.~(\ref{quasiD0}).

In the same way, for the persitence probability, Eq.~(\ref{persl}) can be
rewritten as:
\begin{equation}  
{1\over p_{per}}=LR^{1/\omega}\,\int_0^1 R^{-g} dg\,,
\label{averpersquasi}
\end{equation}  
which is evaluated in~(\ref{quasipers}) using the correspondence between
$\delta$ and $L^{-1}$. 

ii) For the Thue-Morse sequence, the average can be simply performed 
by noticing that $f_{2i-1}+f_{2i}=1$, $01$ and $10$ appearing with the 
same probability , thus $g_{2i}=0$ with probability $1/2$ and 
$g_{2i+1}=\pm1/2$, each with probability $1/4$. Then considering the 
different parity combinations for $n$ and $i$ in~(\ref{dif}), one obtains
the  expression given in~(\ref{D0TM}). In the same way, (\ref{persl})
leads to~(\ref{persTM}) for the persistence probability.
  
\section{Finite-size scaling calculation of the persistence 
exponent for the period-doubling environment}

For the period-doubling sequence, $f_{2i}=1-f_i$ and $f_{2i+1}=0$, so that:
\begin{equation}  
n_{2i}=n_{2i+1}=\sum_{k=1}^i f_{2k}=i-n_i\,.
\label{n}
\end{equation}  
On a system with size $4L$, one has:
\begin{equation}  
{1\over p_{per}(4L)}=S_{4L}(\epsilon,R)=1+\sum_{i=1}^{4L}\epsilon^{-i}
R^{-n_i}\,,
\label{sum}
\end{equation}  
and, splitting the sum into odd and even parts:
\begin{eqnarray}
&&S_{4L}(\epsilon,R)=\nonumber\\
&&\qquad =1+\epsilon^{-1}+\sum_{i=1}^{2L}\epsilon^{-2i}
R^{-n_{2i}}+\sum_{i=1}^{2L-1}\epsilon^{-2i-1}R^{-n_{2i+1}}\nonumber\\
&&\qquad =1+\epsilon^{-1}+\sum_{i=1}^{2L}(\epsilon^2R)^{-i}
R^{n_i}+\epsilon^{-1}\sum_{i=1}^{2L-1}(\epsilon^2R)^{-i}R^{n_i}\nonumber\\
&&\qquad\simeq(1+\epsilon^{-1})\, S_{2L}(\epsilon^2R,R^{-1})\,.
\label{sum4}
\end{eqnarray} 
Iterating one step further, the last relation leads to:
\begin{equation}  
S_{4L}(\epsilon,R)=(1+\epsilon^{-1})(1+\epsilon^{-2}R^{-1})\,
S_L(\epsilon^4R,R)\,.
\label{fss}
\end{equation}  
At the critical point, $\epsilon_c=R^{-1/3}$, this gives:
\begin{equation}  
S_{4L}(\epsilon_c,R)=(R^{1/6}+R^{-1/6})^2\, S_L(\epsilon_c,R)\,.
\label{fss2}
\end{equation}  
Since $p_{per}\sim \delta^\chi\sim L^{-\chi}$, one immediately
recovers the persistence exponent given in Eq.~(\ref{betapd}).
\end{appendix}

\refrule

\end{multicols}

\begin{references}
\bibitem{rem-fibo}
        The substitution rule used here to generate the Fibonacci sequence
        corresponds to the product of two substitutions with the more usual
        rule $0\to 01$, $1\to 0$. It has the advantage of leading by 
								iteration, starting with 0, to a succession of finite sequences which
        are stable from both ends: 0, 010, 01001010,...
\end{references}
\end{document}